\documentclass[twocolumn,showpacs,preprintnumbers,amsmath,amssymb,floatfix]{revtex4}

\usepackage{graphics}
\usepackage{graphicx}
\usepackage{dcolumn}
\usepackage{bm}
\usepackage{amssymb}
\begin{document}
\preprint{APS/123-QED}

\title{Competition between magnetism and superconductivity in
the organic metal $\kappa$-[BEDT-TTF]$_2$Cu[N(CN)$_2$]Br}

\author{David Fournier}
\author{Mario Poirier}
\author{Kim Doan Truong}
\affiliation{ Regroupement Qu\'eb\'ecois sur les Mat\'eriaux de
Pointe, D\'epartement de Physique, Universit\'e de Sherbrooke,
Sherbrooke, Qu\'ebec,Canada J1K 2R1}

\date{\today}

\begin{abstract}
Ultrasonic velocity and attenuation measurements on the
quasi-two-dimensional organic conductor
$\kappa$-[BEDT-TTF]$_2$Cu[N(CN)$_2$]Br give the first
determination of the coexistence zone of the antiferromagnetic
and superconducting phases which extends deep in the metallic
part of the pressure-temperature phase diagram of these salts.
This zone is identified from a precise control of the ordering in
ethylene end groups which acts as an internal pressure. The
two phases are found to compete whereas superconducting
fluctuations begin to contribute to the attenuation at 15 K,
namely at the onset of magnetic order that is well above the
superconducting transition temperature $T_c$ = 11.9 K. Finally,
the temperature profile of sound attenuation for both
longitudinal and transverse phonon polarizations is found to be
inconsistent with a conventional \textit{s}-wave order parameter.
\end{abstract}

\pacs{74.70.Kn, 74.25.Ha, 74.25.Ld }
\maketitle
The physics of strongly correlated materials in the vicinity of
the Mott transition has motivated much interest over the last
decade\cite{KAGAWA2005}. Vanadium oxides, manganites, frustrated
quantum magnetic insulators, high-$T_c$ and organic
superconductors are examples of such systems where complex orders
and quantum criticality can be found. More specifically, the
coexistence and/or competition of magnetism and superconductivity
near the Mott transition in high-$T_c$ and organic materials
constitute fundamental issues that are widely addressed in the
literature (ex. \cite{MILLER2006}). In this respect the layered
quasi-two-dimensional $\kappa$-[BEDT-TTF]$_2$X compounds are of
great interest. Their generic temperature vs pressure phase
diagram \cite{ITO1996,LEFEBVRE2000,FOURNIER2003,LIMELETTE2003}
reveals that the superconducting (SC) and the antiferromagnetic
insulating phases (AF) are separated by a first order Mott
transition line (MI), where a microscale phase separation takes
place \cite{SASAKI2004}. The first order line persists in the
paramagnetic phase and terminates with a critical point, from
which a pseudogap line emerges under pressure as a precursor of
the superconducting state. $\kappa$-[BEDT-TTF]$_2$X compounds can
thus be considered as prototype systems to study the interplay
between magnetic and superconducting phases in the close
proximity of a first order MI transition

In this work the ultrasonic technique and thermal cycling are used
to present a detailed study of the different phases appearing in
the vicinity of the Mott transition line in the
$\kappa$-[BEDT-TTF]$_2$Cu[N(CN)$_2$]Br compound (noted
$\kappa$-H$_8$-Br hereafter). At ambient pressure,
$\kappa$-H$_8$-Br is located on the metallic side of the MI line
and a slight shift on the pressure scale can be achieved by a fine
thermal tuning of residual intrinsic disorder related to ordering
of ethylene end groups of BEDT-TTF molecules occurring between 200
and 60 K \cite{TANATAR1999}. Although a larger shift could be
obtained by partial alloying with the Cu[N(CN)$_2$]Cl anion and/or
from deuteration of the BEDT-TTF molecules, these procedures
introduce, however, extrinsic disorder that must be avoided. Here,
we report sound velocity and the first ultrasonic attenuation
measurements to be realized on these layered compounds. The
results clearly establish a wide domain of the P-T phase diagram
where magnetic and superconducting ordered phases overlap. A fine
tuning of ethylene disorder through thermal cycles is found to
move the system on the pressure scale. The MI is thus reached from
the metallic side where the emergence of an ordered magnetic phase
is found, which is detrimental to superconductivity. The magnetic
phase that set in around 15 K in $\kappa$-H$_8$-Br scales with the
amplitude of the pseudogap anomaly above 30 K or
so\cite{FOURNIER2003}. As for the superconducting transition
occurring at 11.9 K, fluctuation effects are found few degrees
above the critical temperature.
\begin{figure}[H,h]
\includegraphics[width=8.5cm]{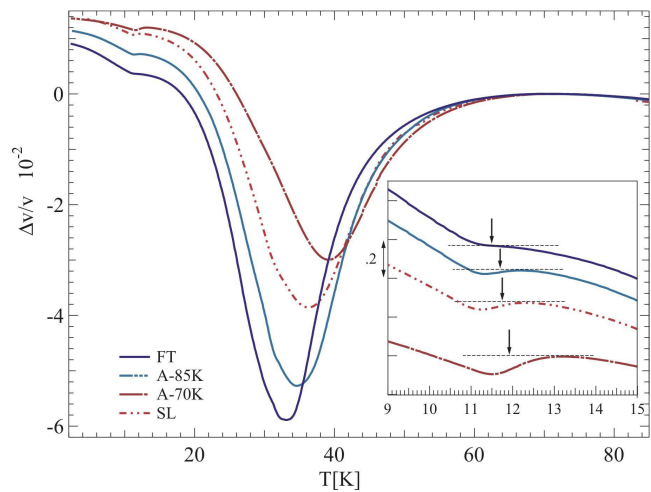}
\caption{\label{fig:fig1}Relative variation of the longitudinal
L[010]velocity at 160 MHz as a function of temperature for
different thermal cycles. Inset: zoom around the 12 K anomaly;
the curves have been shifted for a better view. The arrows
indicate the maximum slope which determines $T_c$.}
\end{figure}

Although difficult to adapt to thin organic crystals, pulsed
ultrasonic velocity experiments have already been carried out on
few compounds of this family
\cite{FRIKACH1999,FRIKACH2000,FOURNIER2003}. It consists in
measuring the phase shift and the amplitude of the first elastic
pulse transmitted through the crystal and a delay line
\cite{FRIKACH2000}; this is why the technique yields only velocity
and attenuation variations. Because of the platelet shape of the
selected $\kappa$-H$_8$-Br crystals, only elastic waves
propagating perpendicularly to the plane (direction (010)) could
be generated with small piezoelectric transducers. Longitudinal
waves polarized along the \textit{b} axis L[010] and, for the
first time, transverse waves polarized along the \textit{a} axis
T[100] could be generated over the 30-450 MHz frequency range. The
crystals were submitted to the following thermal cycles from 200 K
to 2 K: the standard slow cool (SL) process, 10 K/hour, and a fast
cool (FT) process during 10 minutes; following a FT process, a 24
hour annealing at 85 K (A-85K) or 70 K (A-70K). According to the
literature \cite{TANATAR1999,PINTERIC2002,MULLER2002}, maximum
order of the ethylene end groups is obtained when annealing is
performed at a temperature between 60 and 80 K for a period of 24
hours. All the ultrasonic data presented in this paper were
obtained when sweeping the temperature up from 2 K. A magnetic
field up to 16 Tesla could be applied perpendicularly to the plane
in order to completely eliminate the superconducting state.

We report in Figure 1 the temperature profile of the relative
velocity variation ($\Delta v \over v$) of the L[010] mode at 160
MHz over the 2-80K range for different thermal cycles. No
frequency dependence is observed over the 30-450 MHz range; only
the signal to noise ratio increases with frequency. The profiles
reveal two distinct anomalies: i) a huge dip (3-6\%) just above 30
K previously associated to a compressibility increase driven by
the magnetic fluctuations of the electronic degrees of freedom at
the pseudogap \cite{FOURNIER2003}; ii) a much smaller anomaly
around 12 K likely related to the onset of superconductivity. Both
the amplitude and the position of the compressibility anomaly are
highly dependent on the amount of disorder: FT (maximum disorder)
yields the maximum amplitude (6\%) at the lowest temperature (33
K), whereas SL followed by annealing at 70 K (minimum disorder)
reduces by half the anomaly and shifts it to higher temperature
(39 K). Other cooling rates give intermediate disorder and in turn
intermediate results. The process of ordering the ethylene end
groups thus mimics an actual increase of the effective pressure
\cite{FRIKACH2000}. The effect of disorder on the small 12 K
anomaly is shown in the inset of Figure 1. Fast cooling (maximum
disorder) produces a mere change of slope just above 11 K while a
small dip grows progressively with increasing order together with
a small shift to higher temperatures. This dip signals the onset
of the superconducting state in agreement with what is expected
for a longitudinal mode \cite{TESTARDI1975}. However, the increase
of $T_c$ with internal pressure (see inset of Fig.1), confirmed by
SQUID magnetization measurements, is not consistent with
hydrostatic pressure studies. This might be explained by small
residual disorder to which non \textit{s}-wave superconductivity
is particularly sensitive. An important hysteresis is also
observed on the velocity data (also true for the attenuation
presented later) when the temperature is swept up and down between
2K and 20 K, which is consistent with the proximity of the first
order MI transition.

As superconductivity is easily quenched by a field perpendicular
to the highly conducting planes ($H^{\perp}_{c2}$ $\sim$ 12
Tesla), a magnetic field investigation of the low temperature
anomaly appears appropriate to identify the different phases. The
longitudinal L[010] mode and a transverse mode, T[100], which
appears to couple more strongly to the superconducting state, are
used for this investigation. We compare in Figure 2(a) the ${\Delta
v \over v}(T)$ data at 0 and 16 Tesla for the FT process (maximum
disorder).
\begin{figure}[H,h]
\includegraphics[width=8.6cm]{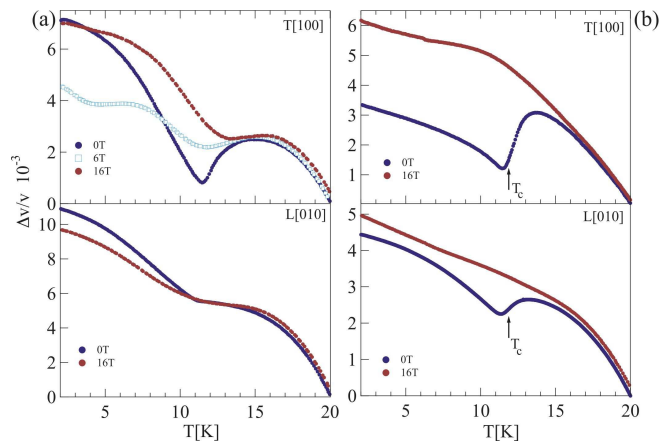}
\caption{\label{fig:fig2}Relative variation of sound velocity as a
function of temperature at fixed magnetic field values: L[010]
mode at 100 MHz and T[100] mode at 160 MHz. (\textbf{a}) FT
process (maximum disorder) and (\textbf{b})A-70K process (maximum
order).}
\end{figure}
 In zero field, the velocity of the L[010] mode shows a
plateau around 15 K before increasing further below 11 K, where a
marked softening of the mode is produced by the magnetic field.
For the T[100] mode, the velocity shows a similar plateau around
15 K, followed by an important dip centered at 11.5 K. The dip
disappears in a 16 Tesla field but the plateau is maintained. We
thus conclude that the 11.5 K dip and the 15 K plateau signal
respectively superconductivity and another ordered phase. This is
confirmed by the 6 Tesla curve of Figure 2(a), which shows the
shift of the 11.5 K anomaly to 5 K, whereas the plateau at 15 K is
kept unchanged. When maximum order (A-70K process) is achieved,
the temperature profile of the anomaly is greatly modified (Fig.
2(b)). When the temperature decreases from 20 K in zero magnetic
field, a maximum is observed around 13.5 K followed by a sharp
drop till a minimum is achieved at 11.5 K; at lower temperatures
the velocity increases smoothly. The amplitude and temperature
profile of the anomaly are however different for both acoustic
modes, being more pronounced for the T[100] one. A 16 Tesla field
completely suppresses the anomaly: the velocity increases smoothly
with decreasing temperature and, in contrast to the FT process, no
plateau is observed at 15 K; only a weak variation of slope occurs
around 8 K. Minimizing disorder then favors the superconducting
phase as signaled by the sharp drop centered at 11.9 K consistent
with the $\Delta v/v$ data (inset Fig.1). We cannot, however,
discard contributions of other phases, which can produce small
slope variations. The presence and nature of theses phases are
most easily revealed by the attenuation measurements whose
sensitivity increases with frequency. We report now the first
ultrasonic attenuation measurements realized on this family of
layered conductors for the A-70K process, which favors
superconductivity.

We present in Figure 3 the variations of the attenuation
$\Delta\alpha (T)$ obtained simultaneously with the ${\Delta v
\over v}(T)$ data. The zeroth value is arbitrarily chosen by
extrapolating the zero field curve at 0 K.
\begin{figure}[H,h]
\includegraphics[width=8cm]{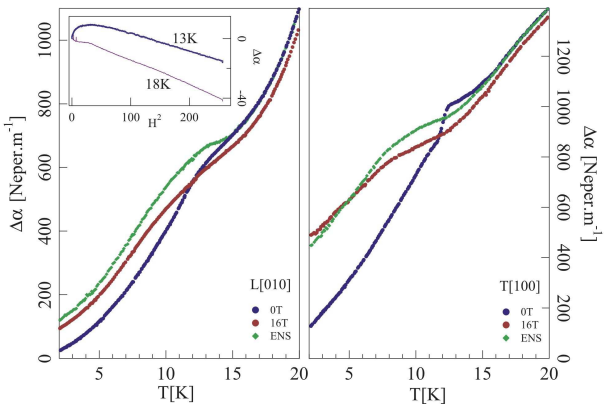}
\caption{\label{fig:fig4}Variation of the sound attenuation as a
function of temperature for the A-70K process (maximum order) at
fixed magnetic field values: L[010] mode at 160 MHz and T[100] at
100 MHz. See text for the description of the ENS curve. Inset:
$\Delta\alpha$ as a function of $H^2$ at 13 K and 18 K for the
L[010] mode at 160MHz.}
\end{figure}
For both ultrasonic modes over the 2-20 K temperature range,
$\Delta\alpha (T)$ is an increasing function of temperature due to
a large attenuation peak accompanying the huge velocity dip near
40 K (Fig.1)\cite{FOURNIER2003}. In zero field, the rate of
increasing attenuation is reduced at the superconducting
temperature around 12 K; this trend is more pronounced for the
T[100] mode, where it almost appears as a discontinuity in the
slope. In a 16 Tesla field, a substantial increase of $\Delta
\alpha$ is produced below $T_c$ when superconductivity is
quenched, while a weak decrease is observed above. These effects
are again more pronounced for the T[100] mode. Surprisingly, a
change of slope persists below 15 K which implies the coexistence
of a secondary phase with superconductivity even when disorder is
minimized. To get an insight into this phase and to isolate the
superconducting contribution to the attenuation, we now examine
precisely the field dependence of the attenuation.

The $\Delta\alpha (T)$ data vary quadratically with field (inset
of Figure 3) over the 2-20 K range as long as $H>H_{c2}^{\perp}$.
Such a $H^2$ dependence, which is also observed for the velocity
data, proceeds from the first correction term to transport and
magnetic properties to which ultrasonic waves couple
\cite{STEINBERG1958,RODRIGUEZ1963,TACHIKI1974,MAEKAWA1977,MAEKAWA1978}.
If we write $\Delta\alpha(T)$ in zero field as a sum of two
contributions, one due to the superconducting state (SC) and
another one to an \textit{exotic normal state} (ENS),
$\Delta\alpha(T)$ = $\Delta\alpha_{SC}(T)$ +
$\Delta\alpha_{ENS}(T)$, we can isolate the SC contribution by
modeling the field effects on the ENS phase in the following way:
$\Delta\alpha_{ENS}(T,H) = \Delta\alpha_{ENS}(T,0) +
\gamma_{\alpha}(T) H^2$. Thus, plots of $\Delta\alpha$ as a
function of $H^2$ at fixed temperatures yield $\gamma_{\alpha}(T)$
as the slope and $\Delta\alpha_{ENS}(T,0)$ as the extrapolated
value toward the zero field limit. The contribution of the SC
phase $\Delta\alpha_{SC}(T)$ is thus obtained after substracting
the ENS contribution.

When compared with zero field data (Figure 3), the
$\Delta\alpha_{ENS}(T,0)$ curves do not appear monotonous. When
the temperature is decreased from 20 K, the ENS curves for both
modes coincide with the zero field one but they depart abruptly
from each other around 15 K. For $T$ $<$ 15 K, apart from the
12-15 K temperature range for the T[100] mode, the ENS attenuation
is higher than in zero field, in agreement with larger sound
attenuation in the normal state compared to the superconducting
one. Anomalous behaviors at 15 K can also be seen in the
$\gamma_{\alpha}(T)$ and $\gamma_v(T)$ coefficients. Examples of
such coefficients for the L[010] mode are presented in the inset
of Figure 4 at a frequency of 428 MHz, which yields maximum
sensitivity. Below 20 K, the $\gamma_{\alpha}(T)$ coefficient
increases with decreasing temperature until an abrupt downward
transition occurs at 15 K; $\gamma_{v}(T)$ is rather constant
below 20 K, but it increases steeply at 15 K with a quasi-linear
variation at lower temperatures. These features of the ENS phase
mimic the onset of an antiferromagnetic order parameter at 15 K
that is 3 K above the superconducting transition. This is
consistent with studies where deuteration of the BEDT-TTF
molecules shifts the compound on the pressure scale at the
boundary of the phase diagram where superconductivity is replaced
by an antiferromagnetic insulating phase at 15 K
\cite{KAWAMOTO1997}. We suggest that the 15 K anomalies observed
in our velocity and attenuation data are the result of a
magneto-elastic coupling with an ordered AF phase. The temperature
at which this phase occurs is $T_{AF}$ = 15 K, and appears to be
very weakly field dependent (Fig. 2a) up to 16 Tesla. This
observation is consistent with the symmetry of the AF phase
deduced for the $\kappa$-D$_8$-Br compound \cite{MIYAGAWA2002}.
\begin{figure}[H,h]
\includegraphics[width=8.5cm]{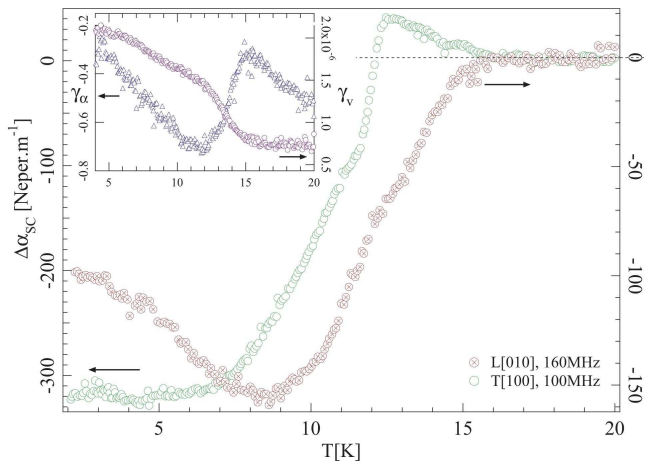}
\caption{\label{fig:fig5}$\Delta\alpha_{SC}$ as a function of
temperature: L[010] mode at 160 MHz and T[100] at 100 MHz. Inset:
Parameters $\gamma_v(T)$ and $\gamma_{\alpha}(T)$ of the ENS
for the L[010] mode at 428MHz.}
\end{figure}

Finally, the contribution of the superconducting state to the
attenuation $\Delta\alpha_{SC}(T)$ is presented in Figure 4. For
both modes, $\Delta\alpha_{SC}(T)$ is negative as expected for a
superconducting state. However, the temperature profile below
$T_c$ does not appear to be consistent with a conventional
\textit{ s}-wave order parameter for which a monotonous decrease
of attenuation is well known to be expected. For the L[010] mode,
the decrease of the attenuation at 15 K suggests the existence of
superconducting fluctuations whereas the shoulder at 11.9 K
coincides with the onset of phase coherence. Then $\Delta\alpha_{SC}(T)$
goes through a minimum around 8 K before increasing further down
to 2 K, while remaining always negative. For the transverse mode
T[100], although superconducting fluctuations are also effective
below 15 K, they rather produce an enhancement of the attenuation
followed by a sharp downward trend below 12.5 K and a further
change of slope at $T_c = 11.9$ K. At lower temperatures
$\Delta\alpha_{SC}(T)$ decreases further and flattens below 7 K.
These features around 7-9 K cannot be reconciled with any
theoretical model for the moment, although we must keep in mind
the peculiar topology of the Fermi surface, which consists of two
parts, namely the quasi-one-dimensional electron sheets and
two-dimensional hole pockets. We remind that the ultrasonic
L[010] mode probes the complete 2D Fermi surface while the
transverse T[100] one is mainly sensitive to the 1D electron
sheets.  Finally, there may be a connection between these low
temperature features around 8 K with a secondary dissipation
phase observed in the pressure transport measurements\cite{ITO1996,ITO2000}.

In summary, our velocity and attenuation data point clearly toward
a competition between metallic (superconducting) and insulating
(antiferromagnetic) phases at low temperatures deep in the Fermi
liquid part of the P-T phase diagram. The farther the system is
from the first order boundary, the stronger is the superconducting
phase relative to the magnetic one. This can be realized by
applying hydrostatic pressure or, as demonstrated in our paper, by
controlling the amount of disorder of the ethylene end groups.
This competition is summarized in Figure 4: when magnetic
fluctuations probably connected to an ordered AF phase disappear
around 15 K, superconducting fluctuations grow rapidly to yield a
transition at 11.9 K. It is clearly demonstrated that disorder
favors the magnetic fluctuations (huge dip at 33 K) which is
detrimental to the superconducting state. However, this
competition apparently yields a phase separation over all the
sample's volume as it was suggested from infrared reflectivity
data at high temperatures \cite{YONEYAMA2005} and confirmed by the
hysteresis observed on our velocity and attenuation data in the
2-20 K range. Finally, although the attenuation data does not
appear consistent with a $\textit{s}$-wave order parameter, a
thorough frequency and polarization analysis of the ultrasonic
attenuation together with precise theoretical predictions
\cite{MARENKO2004} will be needed to establish the symmetry of the
superconducting order parameter.

The authors thank C. Bourbonnais, A.-M. Tremblay for discussions
and the critical reading of the manuscript and M. Castonguay for
his technical support. This work was supported by grants from the
Fonds Qu\'eb\'ecois de la Recherche sur la Nature et les
Technologies (FQRNT) and from the Natural Science and Engineering
Research Council of Canada (NSERC)

\end{document}